\documentclass[twocolumn,showpacs,preprintnumbers,amsmath,amssymb,prl]{revtex4}
\usepackage{graphicx,psfig,rotating,color,psfrag,amssymb,epsfig}% Include figure files
\usepackage{dcolumn}% Align table columns on decimal point
\usepackage{bm}% bold math

\begin{document}

%\preprint{APS/123-QED}

\title{Giant Vortex Lattice Deformations in Rapidly Rotating Bose-Einstein Condensates}\author{T.~P. Simula}\author{A.~A. Penckwitt}\author{R.~J. Ballagh}\affiliation{Physics Department, University of Otago, Dunedin, New Zealand}
\date{\today}
\begin{abstract}
We have performed numerical simulations of giant vortex structures in rapidly rotating Bose-Einstein condensates within the Gross-Pitaevskii formalism. We reproduce the qualitative features, such as oscillation of the giant vortex core area, formation of toroidal density hole, and the precession of giant vortices, observed in the recent experiment [Engels \emph{et.al.}, Phys. Rev. Lett. {\bf 90}, 170405 (2003)]. We provide a mechanism which quantitatively explains the observed core oscillation phenomenon. We demonstrate the clear distinction between the mechanism of atom removal and a repulsive pinning potential in creating giant vortices. In addition, we have been able to simulate the transverse Tkachenko vortex lattice vibrations.
\end{abstract}

\pacs{03.75.LM}% PACS, the Physics and Astronomy
\maketitle

Quantum liquids exhibiting superfluidity are able to support quantized vorticity, as has been often demonstrated in superconductors and helium superfluids. Recently, magnetically trapped gaseous Bose-Einstein condensates (BECs) have been shown to accommodate vortices \cite{Matthews1999b} and arrays of singly quantized vortices \cite{Madison2000a}. In particular, large vortex lattices have been produced by rotating the condensate with an anisotropically deformed trapping potential \cite{Abo-Shaeer2001a} and by cooling the initially rotating thermal gas directly into the vortex lattice state \cite{Haljan2001a}.

Generically, multiquantum vortices (those with a winding number larger than unity) are energetically disfavored and are not expected to persist if created in rotating condensates. However, a number of methods have been studied in order to overcome this vortex dissociation instability, including an external repulsive pinning potential \cite{Simula2002a}. Another, much studied possibility is to use steeper than harmonic trapping potentials to allow rotation of the condensate at frequencies exceeding the centrifugal instability limit of the harmonic trapping \cite{Lundh2002a,Fetter2001a,Fischer2003a,Kavoulakis2002apre,Kasamatsu2002a}. Manifestly multiply quantized vortices can also be created topologically \cite{Isoshima2000a} by deploying the spin degrees of freedom of the condensate accessible in optical traps \cite{Fn1}---a method which has also been realized experimentally \cite{Leanhardt2002a}.

Most recently, giant vortex structures have been produced in rapidly rotating harmonically trapped Bose-Einstein condensates \cite{Engels2003apre}. In this experiment, a near resonant laser beam was shone through the condensate in order to produce a density hole encircled by a high number of vorticity. The mechanism for the removal of the atoms from the beam volume was suggested to be the recoil from spontaneously scattered photons. Using this technique, oscillations of the giant vortex core area and precessing motion of the giant vortices were observed.   

In this paper we simulate the experiment, and compare the effect of atom removal to that of a conservative repulsive pinning potential in the formation and stability of giant vortex structures. Our simulations are in qualitative agreement with the observed experimental results reported in Ref.~\onlinecite{Engels2003apre}, and we provide a simple quantitative explanation for the oscillational behavior of the giant vortex core area. In addition, we have been able to simulate the excitation of the transverse vibrational Tkachenko modes in the vortex lattice \cite{Tkachenko1965a,Anglin2002apre,Baym2003apre} by applying the method used in the recent experiment \cite{Coddington2003apre}.
     
We model the condensate with the usual Gross-Pitaevskii equation
\begin{equation}
i\hbar\frac{\partial \psi({\bf r},t)}{\partial t}= \mathcal{L}({\bf r},t)\psi({\bf r},t)
\end{equation}
for the condensate wavefunction $\psi({\bf r} ,t)$ in a frame rotating with an angular velocity ${\bf\Omega}$. The operator
\begin{equation}
\mathcal{L}=-\frac{\hbar^2}{2m}\nabla^2 + V_{\rm ext}({\bf r},t)+g|\psi({\bf r},t)|^2-{\bf\Omega\cdot L}
\end{equation}
is the Gross-Pitaevskii Hamiltonian, with $g$ determining the strength of the mean-field interactions, and ${\bf L}$ is the angular momentum operator. The external potential $V_{\rm ext}({\bf r},t)$ includes a harmonic trap, 
$V_{\rm trap}({\bf r})=m\omega_\perp^2(x^2+y^2)/2$, and an additional term $V_{\rm laser}({\bf r},t)$ corresponding to the laser beam used to perturb the equilibrium state in the experiment. We model $V_{\rm laser}$ as a localized beam whose time dependence is typically a square pulse, and the spatial shape is given by a Gaussian function of the form
\begin{equation}
V_{\rm beam}(A,x_0,y_0,r_0)=Ae^{-4[(x-x_0)^2+(y-y_0)^2]/r_0^2}.
\end{equation}
Here the amplitude $A$ is either real or pure imaginary, respectively corresponding to a conservative repulsive pinning potential or a localized loss term. The results presented here are obtained, unless otherwise stated, within 2D geometry using $g=1000\,\hbar\omega_\perp\tilde{a}_0^2$, where $\tilde{a}_0=\sqrt{\hbar / 2m\omega_\perp}$. We have initially chosen to normalize the wavefunctions such that the chemical potential $\mu=\langle\mathcal{L}\rangle$ determines the initial particle number.

\begin{figure}[!t]
\includegraphics[width=86mm]{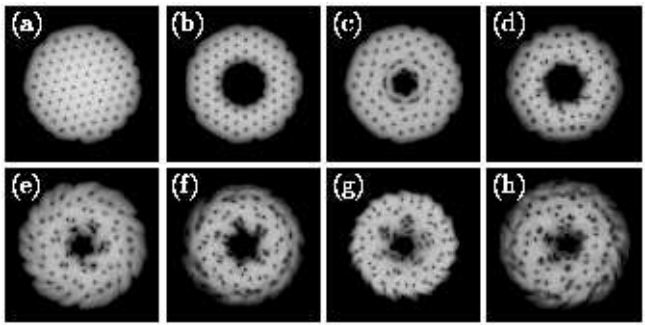}
\caption{Oscillation of the giant vortex core area. The initial equilibrated vortex lattice rotates at angular frequency $\Omega=0.9\;\omega_\perp$, and has $\mu=12\,\hbar\omega_\perp$. Condensate density is locally suppressed from the vicinity of the trap center as described in the text. In effect, the giant vortex core area begins to oscillate followed by the excitation of the overall breathing mode of the condensate. The snapshots (a)-(h) are taken at times $t=(0, 5, 19, 33, 60, 70, 81,$ and $ 100)\times 10^{-2}T$. The sides of each picture are $40\,\tilde{a}_0$ wide.}
\label{Fig1}
\end{figure}

\begin{figure}[!t]
\includegraphics[width=86mm]{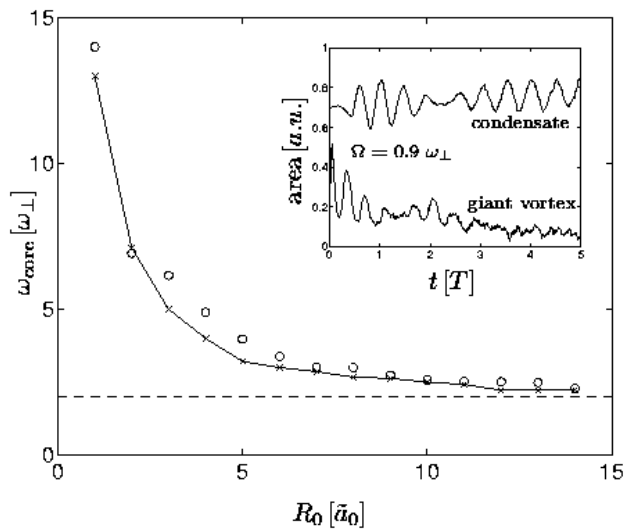}
\caption{Giant vortex core area oscillation frequency as a function of the radius of the giant vortex. Data from our simulations $(\times)$, joined by a solid line for clarity, are plotted together with the predictions $(\circ)$ from Eq.~(\ref{osc}). The dashed line is the breathing mode frequency $2\,\hbar\omega_\perp$ for the 2D system. The data is obtained using the condensate shown in Fig.~\ref{Fig5}(a) as the initial condition. The inset is described in the text, and the lower curve is scaled by a factor of $5$.}
\label{Fig6}
\end{figure}

We start with an equilibrated vortex lattice rotating at a frequency $\Omega=0.9\;\omega_\perp$. An atom-removing loss term, $V_{\rm beam}(-100i\,\hbar\omega_\perp, 0, 0, 7\,\tilde{a}_0)$, is then applied at the center of the trap for the interval, $t=[0,0.05\,T]$, where  $T= 2\pi/\omega_\perp$. The subsequent evolution of the condensate density is displayed in Figs.~\ref{Fig1}(a)-(h) in the laboratory frame of reference. 
The area of the hole begins to oscillate with a dominant Fourier frequency $3.0\;\omega_\perp$ during $t=[0.05,5]\;T$, as illustrated by the lower oscillation pattern in the inset of Fig.~\ref{Fig6} \cite{Fn2}. The contraction and expansion of the hole reflects the hexagonal lattice structure, as is most clearly seen in Fig.~\ref{Fig1}(d). The individual vortex phase singularities within the giant vortices remain separated at all times in contrast to genuine multiply quantized vortex states. The loss of atoms from the trap center also excites the collective breathing mode(s) of the condensate mainly at the universal value $2.0\;\omega_\perp$ \cite{Pitaevskii1997a}, which is also obtained from our calculations for the Bogoliubov modes for the equilibrium vortex lattice eigenstates. In addition, both the core and breathing mode oscillations contain a second order $\Omega$-dependent breathing mode component at $2.4\;\omega_\perp$, which provides coupling between those modes.  

\begin{figure}[!h]
\includegraphics[width=86mm]{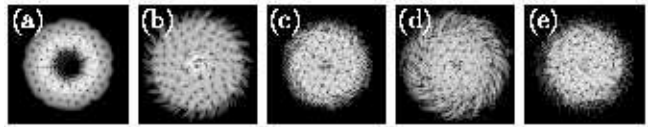}
\caption{Effect of applying a repulsive potential in the center of the vortex lattice, as described in the text. Frames (a)-(e) are for times $t= (11, 27 ,53, 71,$ and $100)\times 10^{-2} T$. The initial configuration is as for Fig.~(\ref{Fig1}). The sides of each picture are $40\;\tilde{a}_0$ wide.}
\label{Fig2}
\end{figure}

The core area oscillation of the giant vortex may be understood in terms of the centrifugal potential associated with the azimuthal condensate flow around the equilibrium radius of the vortex. The condensate flow field around the boundary of the giant vortex may be approximated to be $v=\frac{\hbar}{m}\frac{\ell}{r}$, where $\ell$ is the number of circulation quanta enclosed in the giant vortex of radius $r$. Expanding the centrifugal potential $\frac{\hbar^2}{2m}\frac{\ell^2}{r^2}$ in a Taylor series around an equilibrium radius $R_0$ of the giant vortex, we find the quadratic term in the radial displacement to have the effective harmonic frequency
\begin{equation}
\omega_{\rm core}=2\sqrt{3} \,\ell\left(\frac{\tilde{a}_0}{R_0}\right)^2\omega_\perp,
\label{osc}
\end{equation} 
suggesting the vortex density $\ell/R_0^2$ to be the dominant factor determining the variation of giant vortex core oscillation frequencies.

To test the validity of this formula, we plot in Fig.~\ref{Fig6} our simulation results for the core area oscillation frequency as a function of the giant vortex core radius. The crosses (joined by a solid line to guide the eye) are the oscillation frequencies obtained from the interval between the two first minima in the core size. The circles are the predictions of the Eq.~(\ref{osc}), in which we use for the radius $R_0$ the average of the first maximum and minimum core radii of the giant vortex. The number of circulation quanta $\ell$ is taken to be the number of vortices initially residing within the radius $r_0$ of the beam used to create the giant vortex \cite{Fn3}. For smaller values of $R_0$ (larger $\ell/R_0^2$) the core oscillates rapidly, but with increasing core size, the oscillation frequency slows, and approaches the value of the breathing mode, as the size of the giant vortex approaches that of the condensate itself. We have also performed 3D simulations for $\Omega=0.95\;\omega_\perp$ with parameters corresponding to the JILA experiment \cite{Engels2003apre}, and obtained an initial core oscillation frequency of $2.9\;\omega_\perp$ in excellent agreement with the experiment \cite{Fetter2003apre}. 

Equation (\ref{osc}) also suggests that smaller rotation frequencies lead to higher core oscillation frequencies, since the equilibrium radius $R_0$ of the giant vortex with constant vorticity is proportional to the healing length $\xi\propto (\omega_\perp^2-\Omega^2)^{-1/5}$ \cite{Fischer2003a}. However, in the light of our simulations, a more probable reason for the experimentally observed increase in the core oscillation frequency with decreasing rotation frequency is that the consequent increase in the vortex lattice spacing yielded giant vortices with smaller number of circulation quanta in them (assuming constant radius for the applied atom removing laser beam). For sustained oscillation, it is essential that there exists an energy barrier associated with the single vortices within the giant vortex `re-entering' the condensate. This holds the circulation quanta within the giant vortex together during its oscillatory motion. Eventually, the condensate pushes back into the giant vortex core, and as we see in Fig.~\ref{Fig1}, the first six vortices are lost back into the condensate from the corners of the hexagon at a time between the frames (d) and (e). 
\begin{figure}[!t]
\includegraphics[width=86mm]{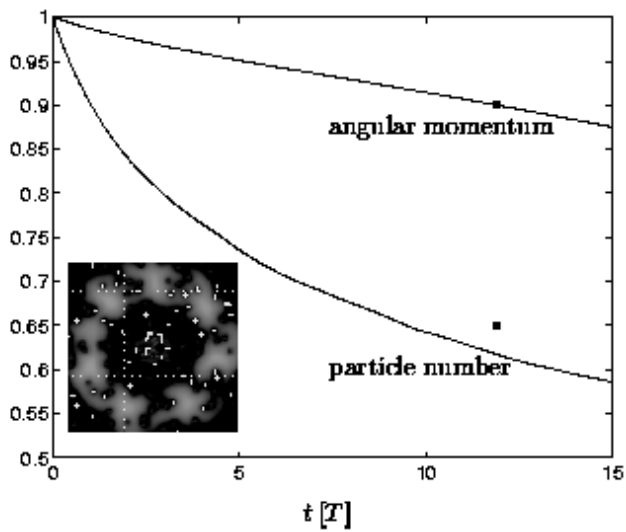}
\caption{Angular momentum (upper curve) and condensate atom number (lower curve), as functions of time, normalized to their initial values. A weak atom removing loss term, $V_{\rm beam}(-0.1i\,\hbar\omega_\perp, 0, 0, 5\,\tilde{a}_0)$, is constantly applied in the center of the condensate rotating at the frequency $\Omega=0.95\;\omega_\perp$. The solid squares are plotted for comparison with the experimental result, see Ref.~\onlinecite{Engels2003apre}. The subplot shows a typical shape of the condensate after it has been strongly depleted.}
\label{Fig3}
\end{figure}
If, instead of removing the atoms from the trap center, we apply a repulsive potential $V_{\rm beam}(100\,\hbar\omega_\perp,0,0,7\,\tilde{a}_0)$, then, as shown in Fig.~\ref{Fig2}, no oscillation occurs in the giant vortex core area. Instead, the individual vortex cores are initially repelled from the trap center together with the condensate and depending on the amplitude of the beam used, shock-wave-like wave fronts and fine interference fringes may be produced. After removing the pinning potential, the mean-field pressure quickly closes the central hole leaving the whole condensate in the breathing mode oscillation. 

\begin{figure}[!t]
\includegraphics[width=86mm]{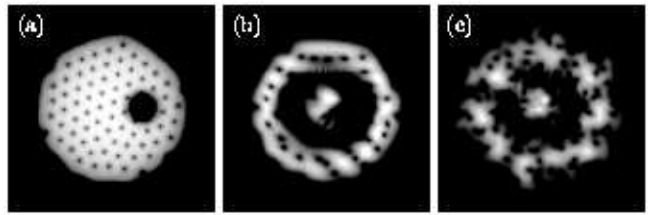}
\caption{Formation of a toroidal hole in the rotating condensate. Localized off-center Gaussian beam suppressing the condensate density is kept still in the laboratory frame. As the vortex lattice rotates, a ring is cut in the condensate. After one trap cycle $T$ the loss term is switched off, after which the ring persists in the condensate for several condensate rotation periods. Frames from left to right are taken at times $t=(0.05, 1,$ and $5$) $T$.}
\label{Fig4}
\end{figure}

These simulations confirm that the mechanism for creation of giant vortices in Ref.~\onlinecite{Engels2003apre} depends in an essential way on the removal of atoms within the volume of the applied laser, and is different from the pinning effect from a repulsive laser beam. The reason is that while the pinning potential conserves angular momentum $L$ and atom number $N$, the absorbing beam {\emph{increases}} $L/N$. This can be understood by noting that for the rotating lattice, the angular momentum $\propto \Omega R^2$ within a circle of radius $R$, so that removal of atoms within $R$ leaves an annulus with higher $L/N$. This annulus can now begin to relax towards an equilibrium vortex lattice, but in the absence of any thermal cloud, it must conserve the value of $L$, since the potential is cylindrical. It is straightforward to show that in comparison to the original lattice, a new lattice with less atoms and higher $L/N$ must have larger angular velocity $\Omega$. Thus as the annulus spreads into the central hole, driven by mean field pressure and confining trap, it encounters an angular momentum barrier because the condensate can not speed up rotationally as fast as the hole is filling. The core area thus begins to undergo oscillation. For the pinning potential on the other hand, even though atoms are repelled from the central region, the condensate retains the appropriate $L$ and $N$ for a vortex lattice rotating at the angular velocity $\Omega$. Hence, when the repelling potential is removed, the core fills immediately with no oscillation. The distinction between the two mechanisms is further emphasized in Fig.~\ref{Fig3}, where we plot the time evolution of the angular momentum of the condensate and the number of particles in the condensate while a weak atom removing loss term is constantly applied. The ratio $10/35$ between the lost fractions of angular momentum and particles reported in the experiment is in fair agreement with the data of our numerical simulations (at the time when 10\% of atoms have been lost). Changing the amplitude $A$ of the loss in the scenario of Fig.~(\ref{Fig3}), merely rescales the time. Furthermore, the shape of the curves shown are only weakly dependent on $g$. The subplot displays how the remaining parts of the condensate tend to break into a number of blobs after the condensate has been intensively suppressed for a long time. Notice, especially, how the individual vortex phase singularities ($+$) are distributed over the whole giant vortex core area. We find in these simulations that the centermost region (the low density area), also contains a number of anti-vortices ($-$), which may be due to numerical noise. Moreover, it is worth pointing out that, unlike an application of a pinning potential, the method of atom removal cannot lead to stable giant vortex configurations as all the atoms in the condensate would eventually be consumed by the atom removing beam.

In Fig.~\ref{Fig4}(a)-(c) we display three frames of the rotating vortex lattice when a localised off-centered atom-removing loss term $V_{\rm beam}(100\,\hbar\omega_\perp, 6\,\tilde{a}_0, 0, 4\,\tilde{a}_0)$ is applied. After one trap rotation period, Fig.~\ref{Fig4}(b), the beam, stationary in the lab frame, has gouged a ring-shaped hole in the condensate. Depending on the radius of the applied loss term this giant vortex ring may persist for several condensate rotation periods. In a similar manner, if the beam is kept on only for a short period of time, the resulting giant vortex begins to precess around the trap center along with the rest of the vortex lattice. Again, if a pinning potential is used instead of removing atoms, neither dark rings nor precessing giant vortices are formed since the suppressed density heals soon after the influence of the pinning potential vanishes. 

\begin{figure}[!t]
\includegraphics[width=86mm]{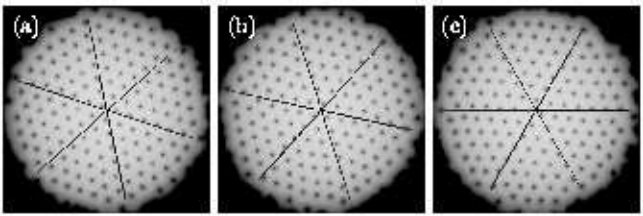}
\caption{Fundamental transverse Tkachenko vortex lattice oscillation mode excited by a localized loss of atoms. (a) The initial vortex lattice configuration with $\Omega=0.95\,\omega_\perp$ and $\mu=11\,\hbar\omega_\perp$ is disturbed by applying a weak atom-removing loss term, $V_{\rm beam}(-0.02i\,\hbar\omega_\perp, 0, 0, 4\,\tilde{a}_0)$, for one trap period $T$. Consequently, the vortex lattice is set into slow vibrational motion around the rotation axis. Figures (a)-(c) are at times $t=(0, 3,$ and $9$) $T$.} 
\label{Fig5}
\end{figure}
We have also performed simulations using various values for the damping term $\gamma$ in a dissipative version of the Gross-Pitaevskii equation, which models the effect of a co-rotating thermal cloud \cite{Fudge2002a,Penckwitt2002a}. Using the damped model does not change the qualitative results here presented but the time scales change and the giant density perturbations tend to heal more quickly. 

Finally, we have studied the recently experimentally observed \cite{Coddington2003apre} transverse Tkachenko oscillations in the vortex lattice \cite{Tkachenko1965a,Anglin2002apre,Baym2003apre}. By suppressing the condensate wavefunction in the trap center with a localized loss term, we are able to induce transverse oscillations in the vortex lattice structure, as demonstrated in Fig.~\ref{Fig5} \cite{BaksmatyMitsushima}. 

In conclusion, we have simulated the formation of giant vortex structures in rapidly rotating Bose-Einstein condensates, and provided a simple mechanism which quantitatively explains the observed core oscillations. We have shown the clear distinction between the mechanism of atom removal and a conservative repulsive pinning potential in creating giant vortices. 
In the JILA experiment the formation of the giant vortex lagged the duration of the laser pulse, while in our simulations such a lag occurs only when the repulsive pinning potential is used. A possible explanation of the lag in the experiment is that the repulsion caused by the atoms leaving the condensate creates outward pressure in the condensate delaying the giant vortex core formation.

\begin{acknowledgements}
Financial support from V\"ais\"al\"a Foundation, Academy of Finland, and Marsden Fund of New Zealand under Contract No. 02-PVT-004 are acknowledged. 
\end{acknowledgements}


\begin{thebibliography}{90}
\bibitem{Fn1}
For topological reasons, the winding number is restricted to even integers in this method.
\bibitem{Fn3}
The two first data points contain large errorbars due to the strong damping of tiny giant vortices. Also determining the appropriate value of $\ell$ to be used for them in Eq.~(\ref{osc}) is not unique.
\end{thebibliography}
\end{document}